\documentclass[a4paper]{article}

\usepackage[english]{babel}
\usepackage[utf8x]{inputenc}
\usepackage[T1]{fontenc}
\usepackage{mdframed}
\usepackage{algorithm}
\usepackage[a4paper,top=3cm,bottom=2cm,left=3cm,right=3cm,marginparwidth=1.75cm]{geometry}

\usepackage{amsmath}
\usepackage{verbatim}
\usepackage{graphicx}
\usepackage[colorinlistoftodos]{todonotes}
\usepackage[colorlinks=true, allcolors=blue]{hyperref}

\title{The Ethereum Scratch Off Puzzle}
\author{
	Abrahim Ladha\\
	\texttt{abrahimladha@gatech.edu}
	\and
	Sharbani Pandit\\
	\texttt{pandit@gatech.edu}
	\and
	Sanya Ralhan\\
	\texttt{sralhan@gatech.edu}
}

\begin{document}
	\maketitle
	
	\begin{abstract}
		Ethereum represents new innovation in the fields of cryptocurrency which has become relatively stagnate, promising many things, including an entire programming language and development enviroment built into the network. However the current trend is to write implementations and proof of concepts before doing the rigor involved with proving security. Miller's recent thesis is an attempt to remedy this, and we apply his provable security techniques to the algorithm description of CASPER, the new ``proof-of-stake" consensus protocol scheme to be implemented in ethereum. We conclude by stating it satisfies almost all the definitions, except one, leaving room for improvement. 
	\end{abstract}
	
	\section{Introduction}
	A very recent field, cryptocurrencies have made their way into spotlight. Cryptocurrencies are virtual currencies that are not administered by any state or corporate entity, but rather exist solely within a decentralized peer-to-peer network that anyone can join. Bitcoin, the first cryptocurrency, has operated with essentially uninterrupted service and significant growth since its launch in 2009. Since money is involved, security is of paramount concern.As they allow open participation from anonymous users, cryptocurrencies derive their security from the strength of underlying network protocol itself. Although Bitcoin has been an empirical success thus far, it is difficult to reason about its security. Miller\cite{miller} has attempted to remedy this in and provide a provable security approach to this usually implementation heavy field. 
	\par CASPER\cite{casper1} is a security-deposit based economic consensus protocol. This means that nodes, so called “bonded validators”, have to place a security deposit (an action they call “bonding”) in order to serve the consensus by producing blocks. The protocol’s direct control of these security deposits is the primary way in which Casper affects the incentives of validators. If a validator produces anything that Casper considers “invalid”, their deposit are forfeited along with the privilege of participating in the consensus process. The use of security deposits addresses the ``nothing at stake" problem \cite{nas}; which suggests that deviation from the protocol is likely to put you at a disadvantage rather than an advantage. In this project we try to fit Miller's definition of provable security into CASPER.
	
	\section{Motivation and Background}
	\subsection{Consensus Protocol}
	Consensus protocols are a field of research which tries to answer the following question "How can $n$ equal parties come to an agreement for some problem?" This has many applications to networks where there is no central authority, but decisions still need to be made and enforced. We discuss methods of obtaining distributed consensus in this section.
	
	\subsection{Proof of Work}
	A proof-of-work (PoW) scheme is a protocol that is difficult to compute but easy to check. The idea is security is reliant upon real world resources, in this case time and hardware. In practice they are often based on hash functions. (With $H$ a collision resistant hash function, For what $s$, is $H(s) = 0^n$ ? What about $0^{n+1}$? and so on.) The only way to solve these problems requires a search of the message space which takes a lot of computing power and much of the security is derived from the security of the hash function. 
	\subsection{Proof of Stake}
	PoW has many problems. It is estimated that the entire bitcoin network will use as much electricity as Denmark by 2020 \cite{denmark}. Proof-of-stake (PoS) is a protocol in which attempts to fix some of PoW's problems. Mining is no longer done on expensive hardware, but rather 'simulated' through probabilities. The resources expended this time are not cpu cycles, but rather the currency itself. For the case of ethereum, you commit to some amount of your coins as your ``stake" and after certain time (currently decided as four months) you are rewarded with your coins back plus some interest. This model is very similar to keeping your money in a bank for its interest rate or a gambling game with a high expected return rate. Proof of Stake has been called ``non-trivial'' to implement by the ethereum developers. \cite{noteasy} 
	\subsection{Ethereum}
	Ethereum is a new cryptocurrency that differs quite a bit from other coins. Ethereum promises an entire programming language and development enviroment built on top of the security of the blockchain. \cite{eth} 
	
	\section{Scratch Off Puzzles}
	\subsection{Definition}
	A scratch off puzzle as defined in \cite{miller} is a tuple $(d,\underline{t},t_0,\gamma)$ and a set of three algorithms:
	\begin{itemize}
		\item $\mathcal{G}(1^\lambda) \rightarrow \texttt{params}$
		\item Work$(\texttt{puz},m) \rightarrow \texttt{ticket}$
		\item Verify$(\texttt{puz},m,\texttt{ticket}) \rightarrow \{0,1\}$
	\end{itemize}
	$(d,\underline{t},t_0,\gamma)$ each mean difficulty, amount of work per puzzle, initialization overhead of the algorithm, and $\gamma$ is the amount an adversary can have an advantage over an honest worker. The optimal idea is that $\gamma$ is as close to 1 as possible. $\mathcal{G}$ initialized all public parameters. Work takes a puzzle instance \texttt{puz} and a payload $m$ and outputs a ticket instance \texttt{ticket}. Verify takes a puzzle instance, a payload, and a ticket and outputs either 1, or 0. Notice the similiarities between this and the triplet of algorithms common in encryption schemes, namely key generation, encryption, and decryption.
	\subsection{Proofs}
	A scratch off puzzle must satisfy three requirements
	\begin{enumerate}
		\item \textbf{Correctness:} For any $(\texttt{puz},m,\texttt{ticket})$, if Work$_{\texttt{ticket}}(\texttt{puz},m) \neq \perp$ then Verify$(\texttt{ticket},\texttt{puz},m) = 1$
		\item \textbf{Parallel Feasiblity:} The honest Work algorithm can be parallelized without much loss, formally:
		\begin{equation*}
		Pr\begin{bmatrix}
		\texttt{params} \leftarrow G(1^\lambda)\\
		\{\texttt{puz}_i,m_i\}_{i \in [q]} \leftarrow A\\
		\forall i : \texttt{ticket}_i \leftarrow Work_t(puz_i,m_i)\\
		\exists i : Verify(\texttt{puz}_i,m_i,\texttt{ticket}_i) \rightarrow 1
		\end{bmatrix} \geq \zeta(1,qt,2^{-d})  \pm negl(\lambda)
		\end{equation*}
		\item \textbf{$\gamma$-Incompressibility:} The work for solving a puzzle must be ``incompressible" The best possible adversary must not be able to speed up the work faster than a factor of $\gamma$. Formally:
		\begin{equation*}
		Pr\begin{bmatrix}
		\texttt{params} \leftarrow G(1^\lambda)\\
		\{\texttt{puz}_i,m_i,\texttt{ticket}_i\}_{i\in [l]} \leftarrow A^{Work}\\
		all~\{\texttt{puz}_i\}_{i \in [l]}~are~distinct~and\\
		\forall i \in [l] : Verify(\texttt{puz}_i,m_i,\texttt{ticket}_i) = 1~and\\
		(puz_i,m_i) \not \in Q
		\end{bmatrix} \leq \zeta(l,\gamma t,2^{-d}) \pm negl(\lambda)
		\end{equation*}
		With $Q$ being the transcript of queries to the Work oracle by our adversary A. Notice that the property of non-malleability (IND-CCA) is built into this definition. The adversary is allowed to see as many valid puzzle, payload, and ticket instances as he pleases, but is unable to forge his own without doing Work. 
	\end{enumerate}
	We also define the $\zeta$ function from above as:
	\begin{equation*}
	\zeta(l,t,d) = 1 - \sum_i^l {t \choose i} 2^{-di}(1-2^{-d})^{t-i}
	\end{equation*}
	which, informally can be described as "$l$ sucesses after $t$ independent Bernoulli trials each with equal probability $2^{-d}$".
	
	\section{Ethereum as a Scratch Off Puzzle}
	\subsection{Definition}
	Ethereum and CASPER is currently still in the design process, and may undergo revisions, so this description (and subsequent proofs) are based off of a loose definition gathered from various less-than-formal specs \cite{spec1} \cite{spec2}
	\begin{figure}[ht]
		\begin{mdframed}[linewidth=0.5mm]
			The many variables of the specific implementation, such as gas, and stake, may be interpreted as puzzle instances or payloads respectively. They are however, independent of the security of the algorithm, so we wont worry about outfitting them.
			\begin{itemize}
				\item $\mathcal{G}(1^\lambda) \rightarrow \texttt{params}$
				\item $\alpha$-Work$(\texttt{puz},m):$\\
				For each round:\\
				commit and pay to produce a new node.\\
				return signiture of this node as \texttt{ticket}
				\item $\alpha$-Verify(\texttt{puz},m,\texttt{ticket}):\\
				$s \stackrel{\$}{\leftarrow} \{0,1\}$ with some fixed, not necessarily uniform probability\\
				return s
				\item $\beta$-Work$(\texttt{puz},m):$\\
				For each round:\\
				For the set of nodes $N$:
				apply stake to any subset  of $N$ you which is any amounts you can\\
				return the nodes you bet on, and the stakes you applied as \texttt{ticket}
				\item $\beta$-Verify(\texttt{puz},m,\texttt{ticket}):\\
				$s \stackrel{\$}{\leftarrow} \{0,1\}$ with some fixed, not necessarily uniform probability\\
			\end{itemize}
		\end{mdframed}
		\caption{The Ethereum Scratch Off Puzzle Algorithm Definition.}
	\end{figure}
	 \begin{figure}
	 	\centering
	 	\includegraphics[width=1\textwidth]{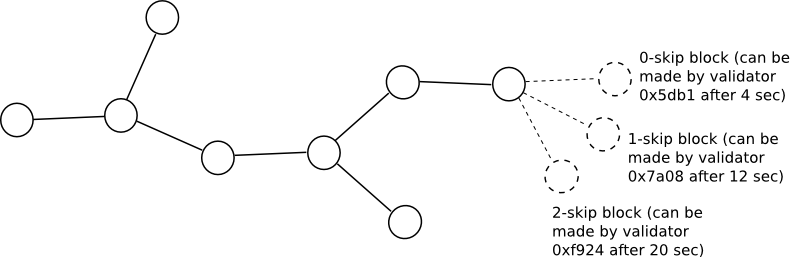}
		 \caption{An example of $\alpha$-Work with three nodes from \cite{spec2}}
	 \end{figure}
	\section{Proofs of CASPER}
	\subsection{Correctness}
	Correctness is trivial in this case since our Verify algorithms are independent of the ticket and return with some probability that is not 1.
	\subsection{$\alpha$-Work $\gamma$-Incompressibility}
	From \cite{spec3} \cite{spec4} we see that  $Pr[\alpha-Verify \rightarrow 1] = \frac{2}{3}$. We also want to conduct only 1 trial, and require it to be successful, therefore $l = t = 1$
	\begin{align*}
	Pr[\alpha-Verify \rightarrow 1] &\leq \zeta(1,\gamma,d)\\
	\frac{2}{3} &\leq \sum_{1}^{1} {\gamma \choose 1}2^{-d}(1-2^{-d})^{\gamma-1}\\
	\frac{2}{3} &\leq \gamma 2^{-d}(1-2^{-d})^{\gamma -1}
	\end{align*} 
	We know that, realistically, $d$ is a super exponential function of time (and its current value is well over $10^{13}$). Analytically evaluating the expression and letting $\gamma$ be a function of $d$, and taking the limit $d \rightarrow \infty$ we see that $\gamma \rightarrow 1$
	
	\subsection{$\alpha$-Work Parallel feasibility}
	\begin{align*}
	Pr[\alpha-Verify \rightarrow 1] &\geq \zeta(1,qt,d)\\
	\frac{2}{3} &\geq \sum_{1}^{1} {qt \choose 1}2^{-d}(1-2^{-d})^{qt-1}\\
	\frac{2}{3} &\geq qt 2^{-d}(1-2^{-d})^{qt -1}
	\end{align*} 
	Letting $q,t$ be reasonable constants for a polynomial adversary and taking the limit $d \rightarrow \infty$ we see the inequality satisfies quite quickly.
	\subsection{$\beta$-Work $\gamma$-Incompressibility}
	There exists a set $N$ of nodes of which an adversary may bet upon. Of these some $S \subset N$ nodes are chosen by the validators, with the nodes belonging to $S$ have the highest bets of any node in $N$. We wish to see if an adversary has some possible advantage. Consider if the adversary uses the greedy algorithm, and places a bet on a single block (denoted $b$) with the greatest current bets. Certainly $b \in S$ if the round were to end immediately. Let $s = |S|$. Then $b$ can be removed from $S$ if $s(bets(b) + 1)$ bets are placed on nodes in $N - S$. Therefore if $k$ is the total amount of bets placed during the round, then $s(bets(b) + 1) > \frac{k}{2} \implies Pr[b \in S] = 1$ by the pigeonhole principle. From \cite{timing} we see that $s \sim 0.4k$. We also want S to vary as a function of $t \in [0,1]$ as time progresses during the round (with 0, 1 being the start and end of the round, respectively). so we let $s = 0.4 \cdot k \cdot t \cdot t_{\texttt{blocktime}}$ where $t_{\texttt{blocktime}}$ is the actual blocktime constant (for bitcoin it is around 10 minutes, for ethereum, it is supposed to be a few seconds). 
	\begin{multline*}
	s(bets(b) + 1) = (bets(b) + 1)\frac{4}{10} \cdot k \cdot t \cdot t_{\texttt{blocktime}} > \frac{k}{2} \implies t\cdot t_{\texttt{blocktime}} > \frac{5}{4(bets(b) + 1)}
	\end{multline*}
	Certainly we see that as the network grows, $bets(b) \rightarrow \infty$ so the only restriction on $t$ is that $t > 0$, so the advantage of betting late diminishes extremely quickly for any sufficiently large network, so we see that $\gamma \rightarrow 1$
	
	\subsection{$\beta$-Work Parallel feasibility}
	The number of validators has been artificially set to a maximum of 8000 (80 shards with 100 validators per shard) \cite{spec2} , as of current discussion. This was raised from an initial proposal of 250 after some concerns.\cite{casper1}. We find that this arbitrary limitation, even with it being as high as 8000 will no less impact the theorectical feasibility. It appears that the inflation is inversely proportional to the number of validators, so the odds of the protocol allowing more than some fixed amount is unlikely. We find that, with this limitation, it is not parallelizable.
	
	\section{Conclusion}
	After analyzing the implementation plans for CASPER and applying the scratch off puzzle security definitions, we have concluded that it satisfies most of the definition, except one.The implementation can be considered secure given if the given scheme can solve this hard cap on the number of validators problem.

\end{document}